Department:
Editor:

# A Real-Time Co-simulation Testbed for EV Charging and Smart Grid Security

**K. Sarieddine, M. A. Sayed**
Concordia University, Security Research Centre

**D. Jafarigiv, R. Atallah**
Hydro-Quebec Research Institute

**M. Debbabi and C. Assi**
Concordia University, Security Research Centre

*Abstract*—**Faced with the threat of climate change, the world is rapidly adopting Electric Vehicles (EVs). The EV ecosystem, however, is vulnerable to cyber-attacks putting it and the power grid at risk. In this article, we present a security-oriented real-time Co-simulation Testbed for the EV ecosystem and the power grid.**

■ THE GROWING ENVIRONMENTAL THREAT has resulted in new governmental policies aimed at combating climate change. Therefore, governments are encouraging the adoption of electric vehicles (EVs) to reduce the emissions of the transportation sector. The global push towards EV adoption has led to exponential growth in EV numbers. To support this rapid growth and improve the quality of charging service, communication networks and Internet-of-Things (IoT) smart charging has been widely adopted in the EV ecosystem. Moreover, to meet the increasing charging demand, operators have been hastily deploying EV charging stations (EVCSs). This has contributed to the ecosystem's lack of proper security measures.

The EV ecosystem is a non-traditional cyber-physical system that has multiple stakeholders operating numerous types of EVCSs which are connected to the power grid. This ecosystem created new attack vectors that could be exploited by adversaries to inflict harm and disrupt power grid operations. The ecosystem has inherited the vulnerabilities of its underlying IoT and communication systems as well as new vulnerabilities originating from the remote capabilities instilled in EVCSs, i.e., communication using Open Charge Point Protocol (OCPP) [1]. Recent events have uncovered real-life attacks against the EV ecosystem. In March and April of 2022, two separate attacks occurred against EVCSs in Russia [2] and the United Kingdom [3] respectively. The attackers in Russia displayed political messages on the EVCSs screens [2] while the attacker in the United Kingdom displayed inappropriate images [3] and rendered the EVCSs unavailable in both cases. An attacker that gained access to do so could easily manipulate the charging profile. Coupled with the fact that attackers can manipulate the charging load itself as demonstrated by the US Department of Transportation [4], EV attacks can have detrimental impacts if used against the grid.

It is imperative to highlight that, due to the interaction of multiple stakeholders (network operators, EVCS manufacturers, and power utilities) the complexity of analyzing the security of the EV ecosystem increases drastically. This work develops and presents a real-time EV co-simulation testbed that provides a realistic environment for security research.

Related Works and Research Gaps

Previous works have implemented digital platforms for evaluation of cyber physical systems performance. We focus mainly on power grid connected technologies. One such example is the work in [5] where the authors survey different digital twin applications in the field of energy storage. They demonstrated how these models can accurately capture the behavior of battery systems, digital model and couples it with digital data to create a digital twin. All these implementations however are only meant to simulate an individual aspect of their respective applications.

The work in [6] on the other hand, creates a digital twin for the integration of blockchain into the field of photovoltaic-connected microgrids. Their work models the power flow and the blockchain using their detailed mathematical models and evaluate the performance of their digital twin in terms of required computing power, energy cost, and grid voltage deviation.

Other studies have attempted to build an EV ecosystem simulator, however, their implementations were restricted to specific components such as the EV, or the scheduling of charging. In [7], the authors propose a design for a real-time simulator of the EV only. The testbed simulates the EV powertrain such as energy consumption monitor, control units, mechanical transmission system, etc. Moreover, details like tire-road and aerodynamic information were taken into consideration. However, their work only focuses on EVs' internal components and does not study the ecosystem as a whole. Our testbed on the other hand focuses on the entire EV ecosystem instead of the EV itself.

In [8], the authors proposed a co-simulator to study the security of the OCPP protocol. Their co-simulator relied on ZeroMQ, Protobus, and a publish-subscribe model to achieve communication between their virtual machines (VMs) hosting their EVCSs, cloud management system (CMS) and the power grid simulator. They used OpenDSS to simulate the distribution power grid. However, this raises the question of scalability of the used grid, as the performance of OpenDSS relies on the computing power of the machine it is running on. Furthermore, their method only focuses on OCPP and disregards the other essential components of the EV ecosystem. On the contrary, our co-simulator includes all the main cyber components of the EV ecosystem and has the ability to simulate the power grid transient and steady state stability in real-time.

Another EV co-simulator was developed in [9] to test a control scheme used to reduce the frequency fluctuations caused by the intermittency of renewable energy resources. The authors develop a control mechanism to store/inject power from the EV batteries into the grid based on frequency fluctuations. Their co-simulator consists of a real-time power grid simulator connected to a power amplifier, which is connected to 2 EVCSs. Although they were able to demonstrate their method using actual EVCSs, their co-simulation was intended to study a single aspect of EV charging and could not be used for security studies since they did not model the actual ecosystem and any of its cyber components and communication channels.

Contributions

Unlike these works that study individual aspects of the ecosystem, we create, to the best of our knowledge, the first co-simulator that includes the different components of the EV ecosystem. So far, we have included the different ecosystem components such as EVCS firmware, realistic EV fleet models, EVCS operator CMS, mobile applications, actual communication protocols, and a real-time power grid simulator. Our testbed provides researchers with the ability to study attacks and their impacts and validate new techniques for securing the ecosystem and the connected infrastructure. We simulate/emulate the different components and their functions and implemented the actual communication protocols used by these systems including our own verified OCPP implementation to communicate between the EVCSs and CMS. We then connect the EVCSs to a real-time power grid simulator. The main elements of this testbed are available on "https://github.com/ksarieddine/A-Real-Time-Co-simulation-Testbed-for-EV-Charging-and-Smart-Grid-Security.git". This co-simulation testbed allows us and other researchers to realistically study the cyber security of the EV ecosystem and its impact on the power grid during attacks and normal behavior.

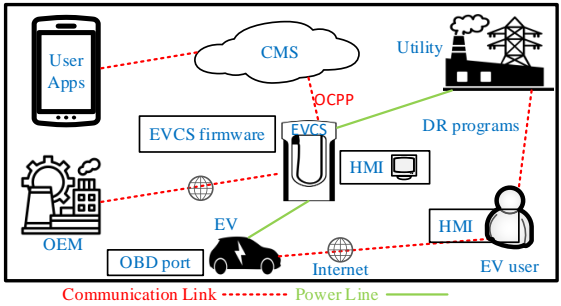

**Figure 1.** Overview of the EV ecosystem

## EV ECOSYSTEM COMPONENTS

As the world is witnessing a great push toward combating climate change, governments have been encouraging the adoption of EVs to reduce emissions from the transportation sector. As such, we are witnessing exponential growth in EV numbers. By the end of 2019, there were 7.2 million EVs on the road [10]. This growth is demonstrated by the record EV sales of 3.2 million in 2020, 6.7 million in 2021, and 10 million in 2022 respectively [10]. This trend is expected to continue for the foreseeable future. Given the large anticipated EV penetration, it becomes imperative that we have a co-simulation testbed that allows us to study all the components of this new ecosystem. Figure 1 depicts the elements of this ecosystem we discuss below.

Overview of the EV Ecosystem Components
- EVCSs: are classified into 3 levels based on their charging rate [11]. Level 1 are slow chargers. Level 2 chargers provide a charging rate of up to 40kW, with 11kW EVCS being the most common. Finally, Level 3 chargers have a charging rate between 40kW to 360kW. While all Level 3 EVCSs are DC chargers, Level 2 can be AC or DC. DC chargers are becoming more common owing to their higher charging rates and ability to support the V2G functionality [11]. EVCSs also host firmware that acts as their local management software. This firmware transforms an EVCS into an IoT device.
- CMS: is a cloud-based software that manages charging sessions and all other public EVCS functionalities using the OCPP protocol [12]. Through this system, users are directed to available EVCS, charging sessions are scheduled and managed and EVCS data is logged.
- User applications: These are phone applications through which users can communicate with the management system over the internet. These services allow users to communicate with the CMS and control charging sessions, pay for charging, control charging rate, and start/terminate charging sessions on public EVCSs.
- EV Protocols: As mentioned earlier, bilateral communication occurs between the components of the EV ecosystem. OCPP is the main protocol used to communicate between the CMS and EVCSs [12].
- Power Grid: The power grid is the backbone of most modern activities and is the source of electricity for the EVCSs.

## METHODOLOGY

This section describes the architecture of our proposed real-time EV co-simulation testbed. Our co-simulator consists of simulated cyber and physical layer components as well as actual communication channels as illustrated in the top-view representation in Figure 2. This testbed can be used for various types of studies related to EV penetration levels, sizing of EVCSs, impact of EV charging on power grid stability, and most importantly evaluating the security of the EV

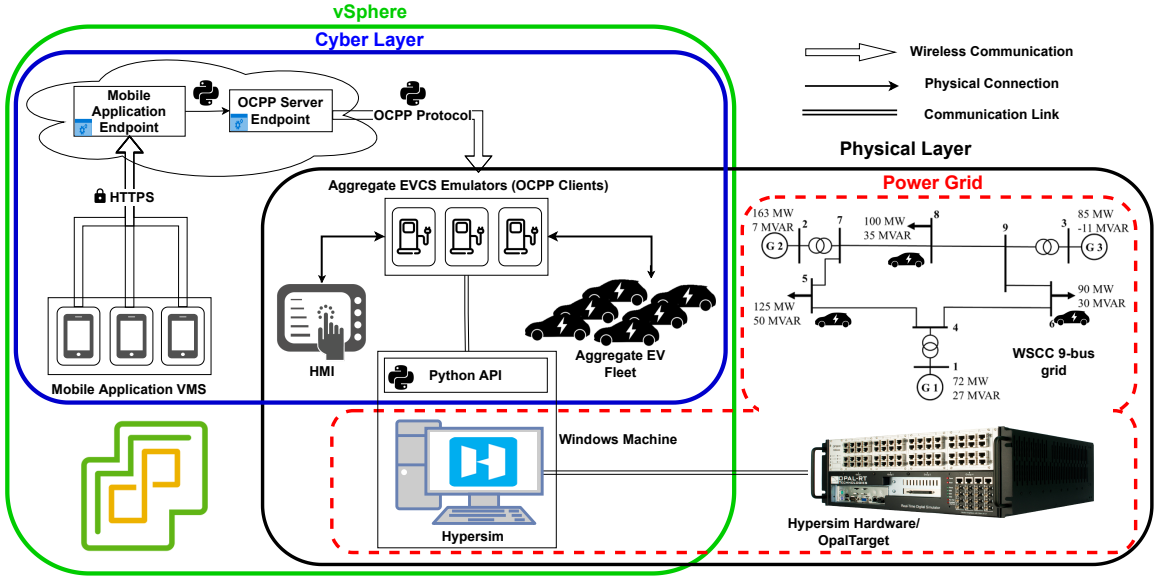

**Figure 2.** Overall co-simulation real-time testbed system model

ecosystem and implications of cyber-attacks on the power grid.

The first step of developing our testbed would be the implementation of the 2 main elements of the ecosystem which are the CMS and the EVCS and establishing communication between them through the de facto OCPP standard. We rely on the basic OCPP setup in [13] and follow the actual OCPP standard [12] to implement the protocol's functions and communication methods. We also implement the phone apps used by users to control their charging sessions and their communication with the CMS. To manage these emulated components and ensure the scalability of our testbed we leverage vSphere which is a VMware cloud computing virtualization platform deployed on a server. vSphere allows us to easily manage extremely large numbers of VMs on which we deploy our CMS, EVCSs, mobile apps, and all other emulated components on Linux-based VMs.

Given the connection of the EVCSs to the power grid, it is of the utmost importance to include the grid in our co-simulation testbed. The EVCSs need to be split among the two layers to be incorporated into our co-simulation testbed. The first layer is the EVCS firmware simulated as part of the cyber layer and the second is the power grid EVCS load.

To simulate the power grid in real-time, we utilize Hypersim which is a power grid simulator by Opal-RT that runs on a windows VM. To achieve real-time simulations, Hypersim runs on dedicated multi-processor hardware (OpalTagert) and is connected to the Hypersim VM over a Local Area Network. We also chose Hypersim for its ability to interact with our emulated components as well as integrated Python scripting abilities.

EV Fleet and EVCS Aggregation

To simulate thousands of users, EVs and EVCSs, we utilized an aggregation mechanism based on the geographic location of the EVCSs. Our aggregation approach is adaptable and could be easily modified based on the studied scenarios and grids. For the sake of demonstration, we aggregate the EVCSs connected at each load bus in our grid into one VM. As for the number of vehicles in our grid, we utilize actual data based on the grid load profile and the number of cars. The specific details are later discussed in the implementation section. While we acknowledge that the current number of EVs is not sufficient to cause large impact on the power grid, the exponentially increasing trend in EV adoption will add a huge EV charging load into the grid as demonstrated in the Experimental Setup section. To this end, after scaling the total number of vehicles to our grid, we perform our study based on a future level of 50% EV adoption. We then determine the number of EVCSs based on the current global EV-to-EVCS ratio and average charging rates. We also create a data-driven model for the arrival and charging times of the EVs. To create a realistic EV load profile we independently simulate a Poisson arrival process of EVs to each EVCS. The charging time of these EVs is then simulated as a truncated Gaussian distribution. The parameters of these models of the arrival and charging time are specified for 1-hour windows for a 24-hour period. These parameters are tuned based on a real dataset containing 5 years of records for 7,000 EVCSs. This dataset was obtained from Hydro-Quebec as part of a legal agreement and research collaboration. Hydro-Quebec owns and operates, through a subsidiary, the public EVCSs in Quebec.

Real-Time Power Grid Simulation

For the real-time modeling of the power grid, we utilize Hypersim which allows us to simulate the power grid with all its static and dynamic behavior. Hypersim provides a flexible and scalable architecture along with high-speed parallel processing to enable real-time and realistic tests to meet the rapidly evolving requirements of the energy sector. Hypersim also allows us to observe, analyze and evaluate the impact of EVCSs on the grid in real-time. Hypersim also includes realistic models of power grid protection mechanisms and the ability to interact with hardware that is connected to the OpalTarget. After we build our power grid model in Hypersim, it is incorporated within our testbed to study the interactions between the grid and the cyber-physical layers of the EV ecosystem. Hypersim also allows us to model any new functionality and control logic we need either by building it directly in Hypersim or importing it from MATLAB Simulink.

To enable the EVCSs VMs to interact with their physical power load on Hypersim, we utilize a Python API that allows the EVCS VMs to control aggregate EV load models on the power grid buses. The delay introduced by the Python script is below 5µs which is negligible. In an actual EVCS, this communication happens over a short wire since the processor sits on top of the hardware making the delays also negligible. To simulate the electric behavior of an EVCS which is a battery charger, we use the PQ-dynamic-load model available in Hypersim. This model allows us to set the loads' voltage sensitivity, frequency sensitivity, harmonics, and all properties of a battery charger.

Cyber and Cyber-Physical Layer Emulation

To create an accurate representation of the EV ecosystem, we first need to emulate the ecosystem's components. As mentioned above, the cyber components are emulated on vSphere. We have chosen vSphere as our platform keeping in mind the need for future scalability as we increase the number of emulated hosts and possibly include different power simulators operating in tandem. vSphere provides us with great flexibility in allocating resources in an efficient manner. The emulated cyber components of the ecosystem include the mobile applications and CMS used to provide remote capabilities and management of the EVCSs, the EVCS firmware and its Human-Machine-Interface (HMI), as well as realistic models of the EV fleet connection to the chargers and their total power drawn from the power grid. Finally, all these components are integrated together by implementing the actual communication protocols they use.

*Mobile application*

The mobile application is an essential component for the commercialization of the EV ecosystem. It provides various remote capabilities such as starting, stopping, and scheduling charging. These mobile applications provide a new attack vector that could be used to impact the power grid [11] by exploiting the lack of end-end authentication between the users and their vehicles. Thus, to study the system comprehensively, we create an emulated version of the mobile application and its functionalities. For the current implementation, we focus on the starting and stopping of charging sessions which we use to demonstrate the attack scenario described below. We create three mobile application VMs that represent the aggregate EV load connected to each bus in the chosen grid. To ensure realistic emulation of the communication initiated by the mobile apps, we use HTTPS communication between the mobile application and the CMS to emulate the real-world communication forced by Android and iOS operating systems. Specifically, we utilize HTTP Post requests to send information to the cloud backend and the HTTP Get request to retrieve information and display it on the mobile app VM.

| Function | From | To | Description |
|---|---|---|---|
| BootNotification | EVCS | CMS | After start-up, a request is sent with information about the EVCS (e.g. version, vendor, serial number, etc.). The receiver then responds to indicate whether it will accept the connection. |
| HeartBeat | EVCS | CMS | A message that ensures availability of the EVCS. |
| Get Configuration | CMS | EVCS | Retrieve the value of configuration settings of the EVCS to understand the available functionalities such as remote charging, smart charging, etc. |
| Authorize | EVCS | CMS | Before the owner of an EV can start or stop charging through the mobile application, the EVCS has to authorize the operation and confirm to the CMS that an EV is connected. |
| Remote Start/Stop Transaction | CMS | EVCS | After authorization, a remote start/stop transactions command can be triggered to initiate the charging session and draw power from the grid (Hypersim in our testbed). |
| Firmware Update and Status | CMS | EVCS | A function that is used to command the EVCS to retrieve an updated firmware from a remote entity whose location is specified via a URL. |
| Charging Profile | CMS | EVCS | a request that contains information about the charging profile including the charging schedule, charger cable limits, etc. |
| Set/Get/Clear Display Message | CMS | EVCS | A set of functions that control the display screen on the HMI and set, retrieve, toggle or clear the messages on the screen. |
| Meter Values | EVCS | CMS | A function that is utilized to send periodic updates to the CMS by collecting readings such as the power consumption, voltage, current, etc. Using these meter values, the CMS gathers granular information about the operation of the EVCSs. |

**Table 1. Implemented OCPP functions and description.**

*Central Management System*

The CMS provides two vital services that are used to enable communication between the mobile app and the EVCS. The CMS possesses a huge amount of computing power and is usually hosted on cloud computing platforms such as AWS, Google Cloud, and Azure. The first service is the mobile application endpoint that is responsible to send and receive requests from the mobile application as discussed above. The CMS will receive start and stop requests to trigger subsequent actions in the cloud where each mobile application VM sends the ID of the EVCS it is targeting. Since we are using aggregated EVs/EVCSs, each of our mobile app VMs also sends the number of EVCSs it is controlling.

Furthermore, the CMS hosts an OCPP server service and has an established communication channel with the EVCSs. The CMS translates the actions triggered by the mobile application to OCPP which is used to manage the EVCSs. The CMS OCPP server is implemented over Python 3.9 following the official standard release [12] and utilized the basic OCPP library [13].

*EV Charging Stations*

The EVCS is the central cyber-physical component of the EV ecosystem. The EVCS would receive requests from the CMS over the OCPP protocol and manages its hardware accordingly. An EVCS is made up of an OCPP client and charging hardware that are common to all EVCSs and firmware that is specific to different manufacturers, as well as an onboard HMI. We implement the OCPP client service on the EVCS to emulate the behavior of a real EVCS and deploy it on a Linux-based VM to represent an aggregate number of EVs/EVCSs following the aggregation logic presented above. The EVCS OCPP client initiates a connection to the CMS to establish a persistent connection that is utilized for all subsequent requests. The OCPP client also keeps track of internal information in a lightweight database, such as EVCS variables that show the status of EVCS (available, busy, error), transaction IDs, etc. To emulate the physical connection of the EVCS to the power grid using the Python API discussed in the Power Grid Simulation Section.

One important function of the EVCS is to verify that an EV is connected and inform the CMS that it can initiate a charging process. Since we are aggregating our EV charging load, the EVCS VM needs to check with the external service hosting the EV fleet model we discussed above for the number of connected EVs at the given bus. Then it reports this aggregate number to the CMS by utilizing the Authorize function implemented over OCPP and described in Table 1. For this work, the EVCS firmware utilizes the Pandas library available in Python to read minute-by-minute EV load values from a CSV file generated as discussed in our EV fleet aggregation section. This will be extended into a digital twin of the EV fleet that changes dynamically in real-time within our testbed.

Finally, we create a representation of the HMI available at the EVCSs that allows local authorization and payment for the charging sessions. This is emulated simply by having a script that can locally initiate a charging session. We also simulate the HMI's display screen by creating the functionality to receive, write and display messages. We aim in our future work to create a visual interface for this display instead of displaying the messages in the console.

*Communication Protocols Implementation*
The OCPP protocol is the de facto standard and the main communication protocol used between the CMS and EVCS. OCPP defines two main roles, a lightweight (client/EVCS) and a central server (server/CMS). OCPP utilizes WebSockets that provides full-duplex communication over a TCP connection. The communication is initiated by the EVCS client when it connects to the CMS server and provides a persistent channel. We have implemented the OCPP protocol and its functions following the official documentation [12]. We implemented the most commonly used version which is OCPP v1.6 and the latest version 2.0.1. It is worth noting that, we validated our OCPP client and server implementations with production-grade EVCSs and the CMS backend provided by Hydro-Quebec as part of a legal agreement and research collaboration.

The communication of the OCPP protocol is in the form of transaction functional blocks, where each entity can initiate a transaction which requires a response from the receiving entity. Initially, when the EVCS is connected to the power outlet, it will send a boot notification to declare its presence to the CMS. The CMS in return replies with the current time, heartbeat interval, and notification if the connection was accepted or rejected. The current CMS time is used to synchronize the clock of the EVCS, whereas the heartbeat interval is used to set the heartbeat frequency of the EVCS which is used by the CMS to validate that the EVCSs are still online. It is worth noting that because of the usage of WebSockets, the OCPP standard mentions that the heartbeat time interval can be as low as once every 24 hours. However, we observed in current practices, heartbeats every 180s. We describe the OCPP functionalities we have implemented in Table 1.

## IMPLEMENTATION AND DEMONSTRATION

The goal of this co-simulation testbed is to generate realistic EV ecosystem behavior for cyber security and power grid stability studies. This testbed can be used to evaluate the security of the EV ecosystem based on actual communication channels, and realistic implementations of the main components within this ecosystem. Our testbed can also be used to collect realistic and real-time data on the power grid's reaction to EV charging load during normal operation and during EV attacks against the grid. Different types of data and communication traffic can be monitored and studied using our testbed ranging from OCPP traffic, mobile applications to CMS communication, EVCS interaction with the power grid, EVCS logs, CMS logs, etc. Using the Hypersim functionalities related to power grid real-time monitoring, we can monitor and record measurements such as voltage and current values, power flows, frequency fluctuations, transformer loading, etc. These measurements can be collected and logged in CSV files to be analyzed later. These measurements can also be incorporated into grid protection mechanisms to add resilience to the power grid whether these mechanisms exist in Hypersim or are implemented by us.

*Experimental Setup and Parameters*
After preparing our testbed, with the previously mentioned emulated components, we generate multiple instances of the EVCSs and mobile apps in order to scale our ecosystem as required. Given our utilization of vSphere, scaling this environment would be rather easy to achieve.

To demonstrate the power grid portion of our co-simulation testbed, we chose to build the WSCC 9-bus grid and implement its detailed and realistic generator models and control mechanisms in Hypersim. The WSCC grid is a sim-plified abstraction of the Western Interconnect in the United States and Canada. This grid has 9 buses, 3 generators, and 3 loads totaling 315MW. For EV fleet aggregation, we consider each of the load buses as a geographical area and aggregate its EVCSs giving us 3 emulated EVCSs and 3 emulated mobile applications. In case of multiple operators, each will have a CMS, 3 EVCSs, and 3 mobile apps. When the setup is complete we scale our EV fleet and our grid loads based on the New South Wales (NSW) grid. To this end, we scale the 9-bus grid based on the NSW load profile [14] and achieve a 24-hour load profile for our testbed. The minimum, average and peak loads in the NSW grid are 5,897MW, 6,968MW, and 8,214MW respectively.

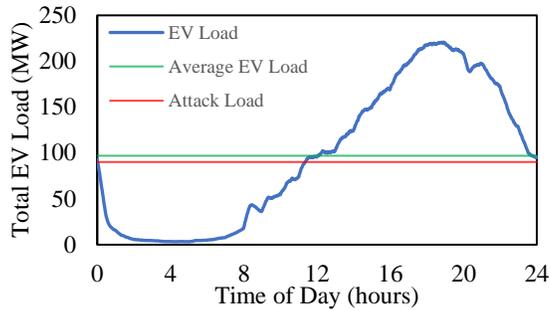

**Figure 3.** Total EV load in 24 hours

The total number of registered vehicles in NSW is 5,892,206 [15]. Scaled down to fit the 9-bus grid, the number of vehicles becomes 266,367. As stated previously, we intend to perform studies on the future impact of EVs on the grid. To this end, we assume a 50% EV penetration level giving us a total of 133,184 EVs in our environment. As per the International Energy Agency (IEA) [10] EVCS operators, utilities, and governments strive to maintain sufficient EVCSs to guarantee the quality of charging services. This has resulted in a global average of 1 public EVCS for every 10 EVs on the road. According to this ratio, our ecosystem will have a total of 13,318 EVCSs distributed proportionally on the load buses. Furthermore, according to the IEA [10], based on the mixture of different charging rates, the global average rate is 24kW per EVCS.

Additionally, we extract the EVCS utilization information from the dataset provided to us by Hydro-Quebec. The average utilization rate of EVCSs is 30-32% with the peak charging demand occurring in the afternoon. By examining the dataset, we extract average hourly arrival rates and charging times and simulate them as a Poisson process and truncated Gaussian distribution respectively. This results in a data-driven model for our EV load for a realistic implementation of our testbed. These details, however, vary from one place to the other where the charging demand in New York, for example, is larger in the morning hours. From the presented statistics and data-driven EV fleet model, we generate the EV load profile presented in Figure 3. Figure 3 demonstrates the minute-by-minute change in the EV load, the daily average EV load, and the magnitude of the EV attack load used below.

*Attacker Model*
We consider a remote adversary that can exploit the vulnerabilities mentioned below to target EVCSs with connected EVs control the EV charging process and coordinate an oscillatory attack against the grid. To demonstrate the operation of our real-time testbed we utilize vulnerabilities discussed in previous work. The EV ecosystem is vulnerable to remote attacks and exploitation by leveraging design flaws [11] and the lack of trust model between the mobile app users and the EVCS they are controlling. The advesrary can then create a botnet of genuine mobile applications to be able to utilize it as an entry point to hijack or initiate an unauthorized charging session remotely [11]. Moreover, the adversary can control ongoing charging sessions by initiating man-in-the-middle (MitM) attacks against OCPP communication using the OCPP vulnerabilities discovered in [1]. Using these vulnerabilities, the adversary can leverage compromised charging sessions to perform large-scale, coordinated attacks against the power grid. We also plan to further develop our EVCS model to include attacks that compromise the firmware of the EVCS itself which depends on specific implementations of different manufacturers.

*EV Attack Scenario*
The EV ecosystem is connected to the power grid, which makes it of the utmost importance to have a realistic testbed to test the EV ecosystem's security and its impact on the power grid. In this paper, we utilize our real-time co-simulation testbed to evaluate the impact of attacks initiated through EV charging loads against the power grid. However, this testbed can be used to study any other types of EV attacks on the power grid.

Oscillatory load attacks are described by an attacker's manipulation of the power grid's load following a certain oscillatory trajectory. These attacks are used to induce forced oscillations on the power grid's frequency making it deviate from the normal 60Hz operating point. This attack is initiated by increasing the power demand to cause a frequency drop and when the system starts its recovery, the attacker would decrease the load to cause a frequency spike. The attack load profile can follow an on/off behavior having a certain periodicity or it can follow a certain periodic waveform such as a sine wave.

In our attack scenario, we initiate the EV load oscillations through the mobile application and OCPP attack models described above. The attacker forces the compromised EVCS to follow an on/off pattern to cause frequency fluctuation on the grid. Our total EV attack load is equal to 90 MW or 3,750 EV/EVCSs split proportionally on the 3 attacked buses. This means that the adversary will have a 12.67-hour window between 11:20 to 24:00 where enough EVs will be connected to the EVCSs. It is noteworthy that other attacks are plausible with smaller compromised EV loads. A smart attacker would target the grid when it is at its weakest point to cause the most damage. Power grids are at their weakest point when their power demand is at its peak in the afternoon or at its lowest point after midnight. Coincidentally, the afternoon is the same time the EV load is at its highest (2.5 times larger than the needed attack load).

For the attack to remain stealthy and hidden from the utility operating the grid, we chose a stealthy/slow oscillatory attack with a frequency of 1 on/off cycle every 5s. Furthermore, this attack remains hidden from EV users since it does not require any change in user behavior. The attacker will only compromise the EVs that are connected to their respective EVCSs at the instance of attack. Furthermore, since an average EVCS would deliver less than 0.9 miles of charge in 1 min (slightly varies depending on the EV properties), an attack lasting a few seconds will have an unnoticeable impact on an EV's range.

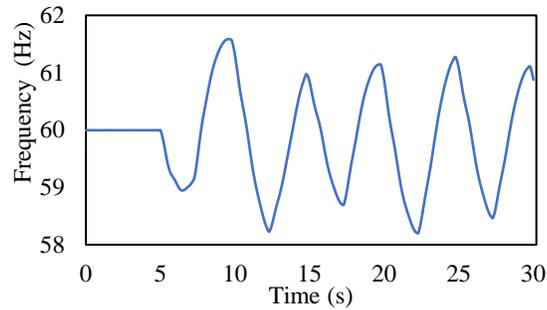

**Figure 4.** Grid frequency response due to EV attack

*EV Attack Impact and Data Collection*

The described attack is initiated and t=5s and will cause the frequency oscillations depicted in Figure 4 on all the grid's buses and its generators while simultaneously causing voltage fluctuations. These frequency oscillations and the deviation beyond 61.5Hz would trip the grid's generator protection relays. The entire grid in this case will lose electricity and enter into a state of blackout. We mention that most utilities have a more stringent requirement and would trip their generators when they experience 1Hz deviations. Even when the attackers control a small number of compromised EVs which is not enough to cause major frequency violations, a sustained attack will hinder the grid from returning to normal operation. A sustained oscillatory load attack would damage the generator turbines due to the induced acceleration and deceleration. Other implications include damaging electric appliances such as EVCSs and home appliances by forcing to operate at frequencies and voltages outside their rated limits.

To visualize this attack, Hypersim offers a functionality called ScopeView that allows us to monitor any measurable parameter in real-time and displays it in a plot format. For more advanced applications, we choose to collect the data from Hypersim ScopeView or directly through a Python API and store them in CSV format to be used for data analytics later on. Figure 4 was generated through such a method where the instantaneous frequency values were stored in a lightweight data storage and used to plot the frequency response of the grid under attack.

## CONCLUSION

In this work, we introduce a real-time co-simulation testbed that emulates the components of the complex EV ecosystem and provides a comprehensive environment to assess its security. The EV ecosystem is complex due to the collaboration of multiple stakeholders (operators, manufacturers, users, utility, etc.). We then use our testbed to demonstrate the impact of EV-based oscillatory attacks against the power grid stability. In future work, we intend to study different cyber-attack vectors in the EV ecosystem. We also aim to utilize our testbed to test different detection, mitigation, and scheduling techniques and study their impact on the power grid. Our testbed is the first to provide such a comprehensive simulation ecosystem in real-time. In the future the EVCS model will be developed to include the specific firmware of different manufacturers. Additionally, an advanced simulation model such as CityMoS (https://citymos.net/) will be used to model the mobility patterns of EVs.

## ACKNOWLEDGMENT

K. Sarieddine and MA. Sayed contributed equally to this paper. This research was conducted as part of the Concordia University/ Hydro-Quebec/ NSERC research collaboration. Grant: ALLRP567144-21.

## ■ REFERENCES

**Khaled Sarieddine,** is a Ph.D. Candidate in Information and Systems Engineering at Concordia University, Canada. He received his BSc. in computer science and his MSc. from the American University of Beirut, Lebanon in 2017 and 2020. His research interests include security of cloud and edge computing, cyber-physical systems, and malware. Contact: khaled.sarieddine@mail.concordia.ca.

**Mohammad Ali Sayed,** is a Ph.D. Candidate in Information and Systems Engineering at Concordia University, Canada. He received his BE in electrical engineering and his MSc from the Lebanese American University, Lebanon in 2015 and 2020. He worked in the operation and maintenance of power plants and his research interests include and power grid stability and security. He is an IEEE Student Member. Contact: mohammad.sayed@mail.concordia.ca.

**Danial Jafarigiv,** is a researcher at the Hydro-Quebec Research Institute. He received his Ph.D. in electrical engineering from the Polytechnique Montreal, Canada, in 2021 and his MSc from the Politecnico di Milano, Italy, in 2015. His current research is on cyber-physical systems, co-simulation platforms, and smart grid cybersecurity. He is an IEEE Member. Contact: jafarigiv.danial2@hydroquebec.com.

**Ribal Atallah,** is a researcher at the Hydro-Québec Research Institute. He received his Ph.D.in information and systems engineering from Concordia University, Canada, in 2017 and his MSc in computer engineering from Lebanese American University in 2012. His current research is on machine learning to protect the smart grid against cyber-attacks. His interests include deep learning, reinforcement learning, and intelligent transportation systems Contact: atallah.ribal@hydroquebec.com.

**Mourad Debbabi,** is a Professor and Dean at the Gina Cody School of Engineering and Computer Science at Concordia University, NSERC/Hydro-Quebec/Thales Senior Industrial Research Chair in Smart Grid Security. He is the founder and the Director of the Security Research Centre at Concordia. He holds Ph.D. and MSc in computer science from Paris-XI Orsay, University, France. He is a co-founder and Executive Director of the National Cybersecurity Consortium. He serves on the expert committee of the Ministry of Cybersecurity and Digital Technology of Quebec. He is an IEEE member. Contact: mourad.debbabi@concordia.ca.

**Chadi Assi,** is a professor and research chair at Concordia University, Canada. He received his PhD from City University of New York where his thesis received the prestigious Mina Rees Dissertation Award. His research interests are include networks, cybersecurity, cyberthreat intelligence, and 5G technologies. He is an IEEE fellow. Contact: chadi.assi@concordia.ca.